\documentclass[namedreferences]{kluwer}
\usepackage[dvips]{epsfig}
\usepackage{graphicx}
\usepackage{amssymb}

\begin{document}
\begin{article}
\begin{opening}

\title{MULTI-WAVELENGTH SIGNATURES OF MAGNETIC RECONNECTION OF A FLARE ASSOCIATED CORONAL MASS EJECTION}

\author{BHUWAN \surname{JOSHI}$^1$, P. K. MANOHARAN$^2$, ASTRID M. VERONIG$^3$, P. PANT$^1$ AND KAVITA PANDEY$^4$}

\institute{$^1$Aryabhatta Research Institute of Observational Sciences,  Manora Peak, Nainital 263 129, Uttaranchal, India (e-mail: bhuwan,ppant@aries.ernet.in)}

\institute{$^2$ Radio Astronomy Centre, Tata Institute of Fundamental Research, Udhagamandalam (Ooty) 643 001, Tamilnadu, India (e-mail: mano@ncra.tifr.res.in)}

\institute{$^3$ IGAM/Institute of Physics, University of Graz, Universit$\ddot{a}$tsplatz 5, A-8010 Graz, Austria (e-mail: asv@igam.uni-graz.at)}

\institute{$^4$ Department of Physics, Kumaun University, Nainital 263 002, India} 

\begin{ao}
Bhuwan Joshi\\
Aryabhatta Research Institute of Observational Sciences (ARIES),\\
Manora Peak, Naini Tal$-$263 129,\\
INDIA.\\
Phone: +91-5942-235583, 235136\\
Fax: +91-5942-233439\\
email: bhuwan@aries.ernet.in
\end{ao}

\begin{abstract}
The evolution of an X2.7 solar flare, that occurred in a complex $\beta\gamma\delta$ magnetic configuration 
region on 2003 November 3 is discussed utilizing a multi-wavelength data set. 
The very first signature of pre-flare coronal activity is observed in radio wavelengths 
as type III burst that occurred several minutes prior to the flare signature in H$\alpha$.
This type III is followed by the appearance of a looptop source in hard X-ray (HXR) 
images obtained from RHESSI. During the main phase of the event, H$\alpha$ images observed 
from the solar tower telescope at ARIES, Nainital, reveal well-defined footpoint (FP) 
and looptop (LT) sources. As the flare evolves, the LT source moves upward and the 
separation between the two FP sources increases. The co-alignment of H$\alpha$ with HXR 
images shows spatial correlation between H$\alpha$ and HXR footpoints, while the rising 
LT source in HXR is always located above the LT source seen in H$\alpha$. The evolution 
of LT and FP sources is consistent with the reconnection models of solar flares. The 
EUV images at 195~{\AA} taken by SOHO/EIT reveal intense emission on the disk at the 
flaring region during the impulsive phase. Further, slow drifting type IV bursts, 
observed at low coronal heights at two time intervals along the flare period, indicate 
rising plasmoids or loop systems. The intense type II radio burst at time in between 
these type IV bursts, but at a relatively larger height indicates the onset of CME 
and its associated coronal shock wave. The study supports the standard CSHKP model of 
flares, which is consistent with nearly all eruptive flare models. More important, 
the results also contain evidence for breakout reconnection before the flare phase.

\end{abstract}

\end{opening}

\section{Introduction}

Stressed magnetic fields are thought to provide the free energy for 
a wide range of transient energetic phenomena on the Sun, ranging
from the smallest microjets and microflares to the largest flares 
and coronal mass ejections (CMEs). During the onset of flares and
CMEs, plasma of a wide range of temperatures (i.e., the cool 
plasma eruption observed in H$\alpha$ to the rapidly heated plasma 
in excess of 10 million K as recorded in X-rays)
as well as fast nonthermal particles, diagnosed in hard X-rays 
and radio, are emitted. 
The multi-wavelength observations are therefore essential to study 
the dynamical phenomena, which include plasma and filament eruptions, 
associated with the various stages of flares and their energy 
generation sites. For example, the radio observations at different 
frequency bands have provided the location, timing, and characteristics 
of beams of accelerated electrons from the flare site (e.g., Manoharan 
et al., 1996). In such multi-wavelength studies, the relationship 
inferred between the hottest thermal electrons and the non-thermal 
electrons provide the particle acceleration sites. However, it is 
unclear how these electrons are accelerated to various energy levels 
and what is the role of triggering of reconnection in controlling the 
timing of the hot thermal and non-thermal emissions and the escape of 
particles into the interplanetary medium.

A simplistic picture of the magnetic configuration of the flaring 
region was revealed by the soft X-ray observations from the Yohkoh 
satellite. It shows a cusp-shaped magnetic field configuration, which 
is formed along a filament channel and consists of overlying
loops system. This simple geometry is in good agreement with the
standard flare models, which demonstrate the reconnection between the 
low-lying current carrying twisted field lines and over-lying casual 
loops and the closure configuration of the system of loops after the flare 
and/or CME 
(see review by Hudson et al., 2004). 
However, recent 
observations and numerical simulations have suggested a more complex 
geometry for the reconnection of the magnetic fields having multiple 
poles (e.g., Antiochos, De Vore, and Klimchuk, 1999; Manoharan and 
Kundu, 2003; Moore et al., 2001). In this model, the energy for the 
flare/CME is stored in the stressed core of multi-polar magnetic fields and
the evolution of their geometry as well as the weakening of the 
surrounding and over-lying fields trigger the onset of rapid magnetic
reconnection to result in an explosion. 

\begin{figure}
\centering
\epsfig{file=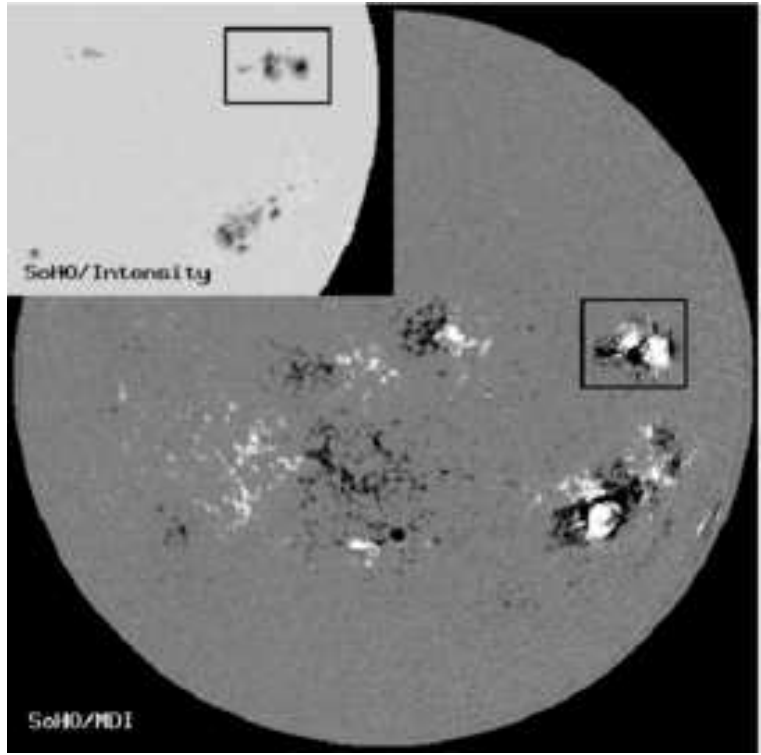,width=10cm, height=10cm}
\caption{SOHO/MDI magnetogram observed
on November 1, 2003, two days prior to the reported event. The position of active region NOAA 10488 is shown inside
the box. The $inset$ shows the SOHO white light image of the active region.}
\label{mdi}
\end{figure}      

In this paper, we present a multi-wavelength study of a flare event (2B/X2.7) that
occurred in the NOAA active region 10488, on November 3, 2003, during
01:00 UT to 02:30 UT. This event also produced a fast coronal mass 
ejection. The near-Sun manifestations of this event have been observed 
in H$\alpha$ (ARIES), hard and soft X-rays (RHESSI and GOES), extreme 
ultra-violet (SOHO/EIT), and radio (HiRAS). In the interplanetary medium, 
the white-light images (SOHO/LASCO), space-based radio spectral data 
(Wind/WAVES), and interplanetary scintillation images (Ooty IPS data)
provide the characteristics of propagation of the CME-associated disturbance.
We describe the observational characteristics of the event 
obtained from the above stated measurements in Section 2. 
We discuss the results and conclude in the final section.

\section{Observations}

The flare event on November 3, 2003 occurred in a complex $\beta\gamma\delta$ magnetic configuration region. 
The SOHO/MDI magnetogram of this active region observed a few days prior to the reported event is 
shown in Figure \ref{mdi}. The evolution of the long duration event (LDE) in H$\alpha$ is shown in
Figure \ref{halpha_plot}, between 01:11 UT and 02:05 UT.

The soft X-ray flux recorded by the GOES satellite in the  0.5--4 and 1--8~{\AA}
wavelength bands starts to steadily increase as early as 00:30~UT, whereas
the fast rise sets in around 01:00~UT and intensifies further at 01:13~UT (cf. Figure \ref{goes_halpha}a).
The maximum of X-ray intensity is
observed around 01:30 UT and then this long duration event 
continues to decline gradually. The important signature observed in
these X-ray profiles is that they indicate the characteristics of
superposition of two peaks. The first one occurs at 01:17 UT and
the later one at 01:30 UT, which coincides with the flare maximum.
For this flare event, as shown by the RHESSI
measurements, the hard X-ray activities seem to peak around the first
phase and in the later time, the soft X-ray flux increases further and
gets intensified at the flare maximum, $\sim$01:30 UT. We will now analyse the
event in detail.

\subsection{H$\alpha$ Images}

\begin{figure}
\centering
\epsfig{file=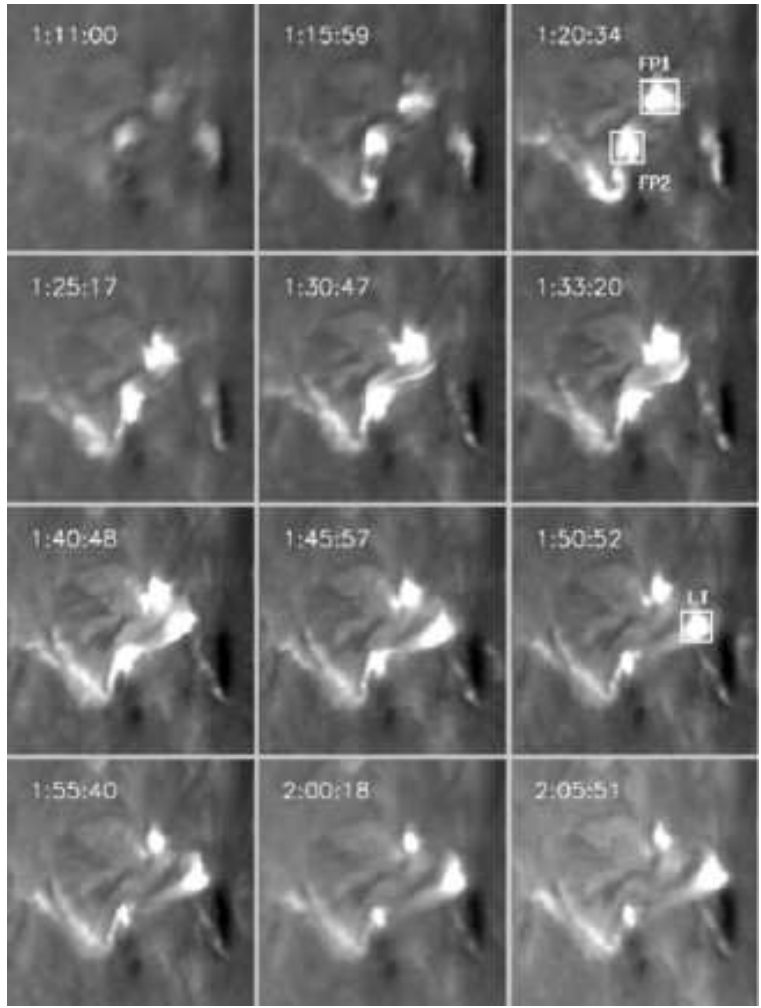,width=10cm}
\caption{Some representative H$\alpha$ images showing the evolution of the long duration flare event from
its rising to declining phase. The size of each image is $100'' \times 100''$. The image at 01:20:34 UT shows
the location of the northern and southern footpoint (FP1 and FP2, respectively). The position of the 
growing looptop (LT) source is marked in the image at 01:50:52~UT.}

\label{halpha_plot}
\end{figure} 

\begin{figure}
\centering
\epsfig{file=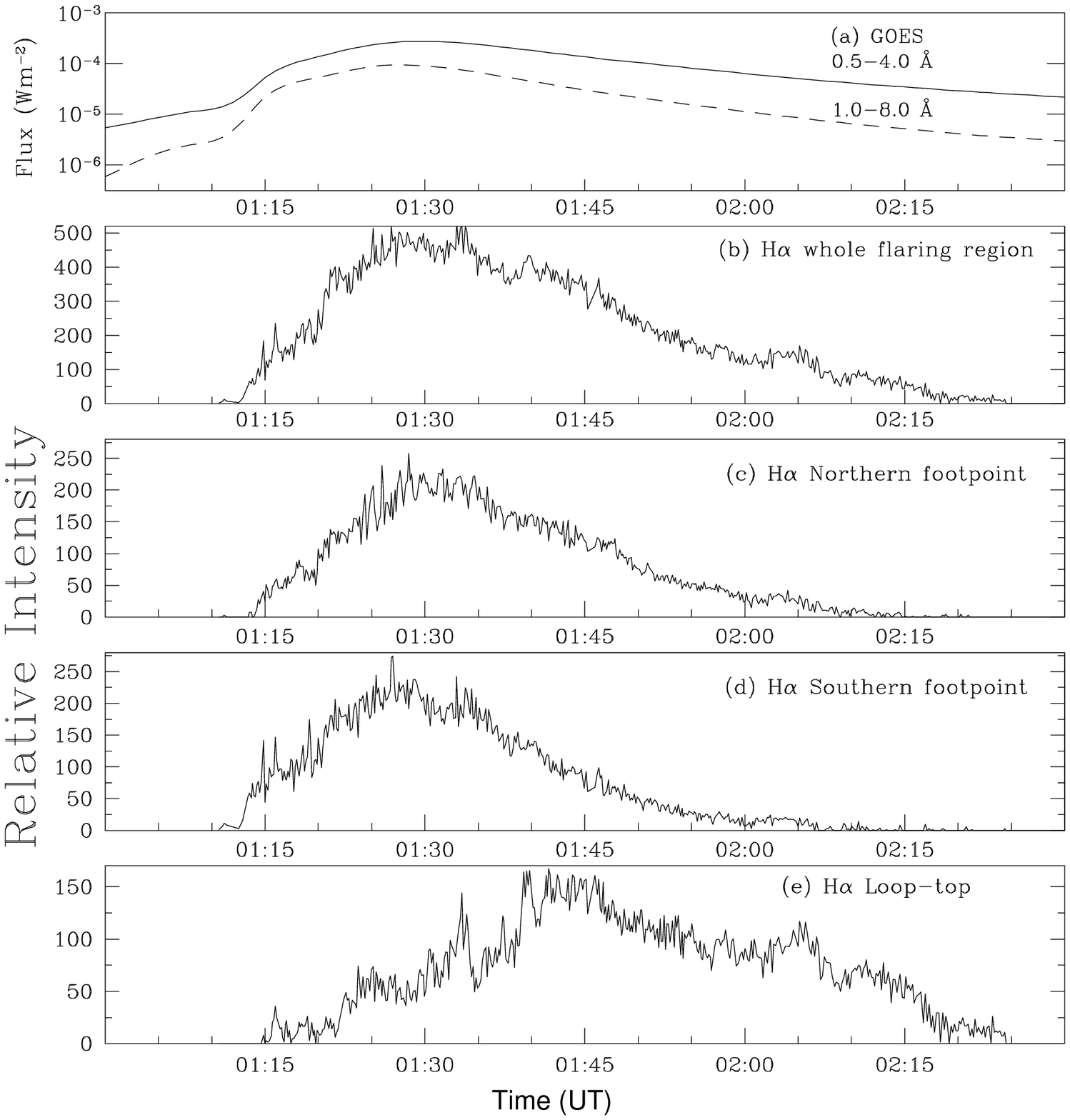, width=12cm}
\caption{(a) GOES X-ray flux in the 0.5--4~{\AA} and 1--8~{\AA} wavelength band. Time profile of the H$\alpha$ intensity 
with respect to the background emission for the whole flaring region, northern footpoint, southern footpoints and 
looptop source (b--e).}
\label{goes_halpha}
\end{figure}

The H$\alpha$ observations ($\lambda$=6563~{\AA} $\pm$ 0.5~{\AA}) of the flare event were made 
using a 15 cm f/15 Coud\'e refractor mounted on a tower at ARIES, Nainital, India. 
The images were recorded with a $385 \times 578$ pixel 16~bit frame transfer 
CCD camera. 
The spatial resolution of the images is about 1$''$ 
per pixel and the temporal cadence is about 20 s. During the 
observation of this event, the seeing condition was favourable and good 
quality images were obtained.

A careful examination of the H$\alpha$ images reveals interesting
features associated with this event as well as with the activity
site. Figure \ref{halpha_plot} shows some representative H$\alpha$ images
from the rising to the declining phase of this long duration event. 
We find that during the rising phase of the flare, there are four chromospheric
sources of H$\alpha$ emission. But only two of them show strong emission (one at the northern and another at 
the southern side of the active region, indicated as FP1 and FP2 in Fig.~\ref{halpha_plot}) and they remain visible
up to the maximum as well as in the decay phase of the flare.
The H$\alpha$ images of these sources after the flare maximum
(i.e., after 01:30 UT) reveal that these two bright regions
are connected by loops.
During the rising phase, a spray-like activity is observed from the southern footpoint 
which shows mass motion towards the south east of the active region.

In Figures \ref{goes_halpha}(b--e), the H$\alpha$ intensity is plotted as a function of time. The 
four panels show the brightness variations of the whole flaring region, the northern footpoint,
the southern footpoint and the LT source with respect to the background emission.
The light curve of the flare as well as
the two footpoints show a fairly quick rise similar to the soft X-ray
profile (cf. top panel of Figure \ref{goes_halpha}) and gradual decline of intensity between 01:10 UT and 02:25 UT.
The brightness of the  
rising looptop source increases after the peak seen in the H$\alpha$ light curve of the 
whole flaring region (cf. bottom panel of Figure~\ref{goes_halpha}).
In the successive H$\alpha$ filtergrams, a rapid increase 
in the looptop brightening is observed as it moves towards the 
west side of the active region. At the same time, the size and intensity 
of the footpoints decrease gradually after the flare maximum and the separation between the
two increases. It is interesting to note that the looptop source in H$\alpha$ is
seen in emission against the solar disk, which is rather an unusual feature. This is considered as evidence
for high pressure/density in the loops (Heinzel and Karlicky, 1987).
Starting from $\sim$01:40 UT, the total intensity of the flaring region seems 
to be dominated by the emission originating from the top of the rising loops.
On the other hand, the overall size and thickness of the loop system 
increase with time and provide a clear picture of the post-flare configuration associated
with the flare event.
 
Here it is relevant to mention that another flare of class X3.9 occurred 
in the same active region nine hours later which looks totally homologous
to this event. Several aspects of this later X-class flare have been 
studied in detail (Liu et al., 2004; Veronig et al., 2006; Vr\v{s}nak et al., 2006a,b; 
Dauphin et al., 2006) 
and they show various common features between the two events in different wavelengths. 

\subsection{RHESSI Measurements}

\begin{figure}
\centering
\includegraphics[width=8cm]{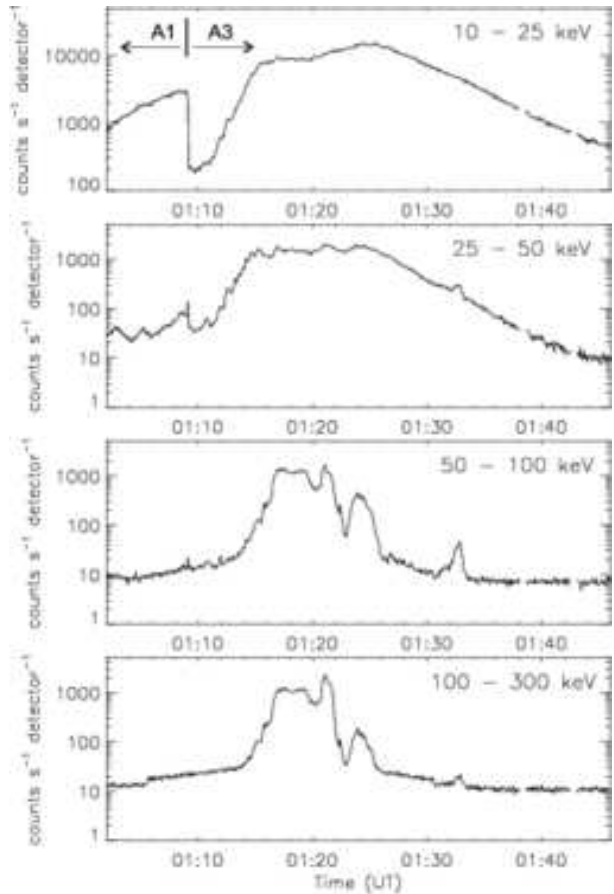}
\caption{RHESSI X-ray lightcurves reconstructed in four energy bands with 4~s integration time.
The change from attenuator state A1 to A3 at 01:09 UT is indicated in the first panel.}
\label{rh_lc} 
\end {figure}

\begin{figure}
\centering
\includegraphics[width=11cm]{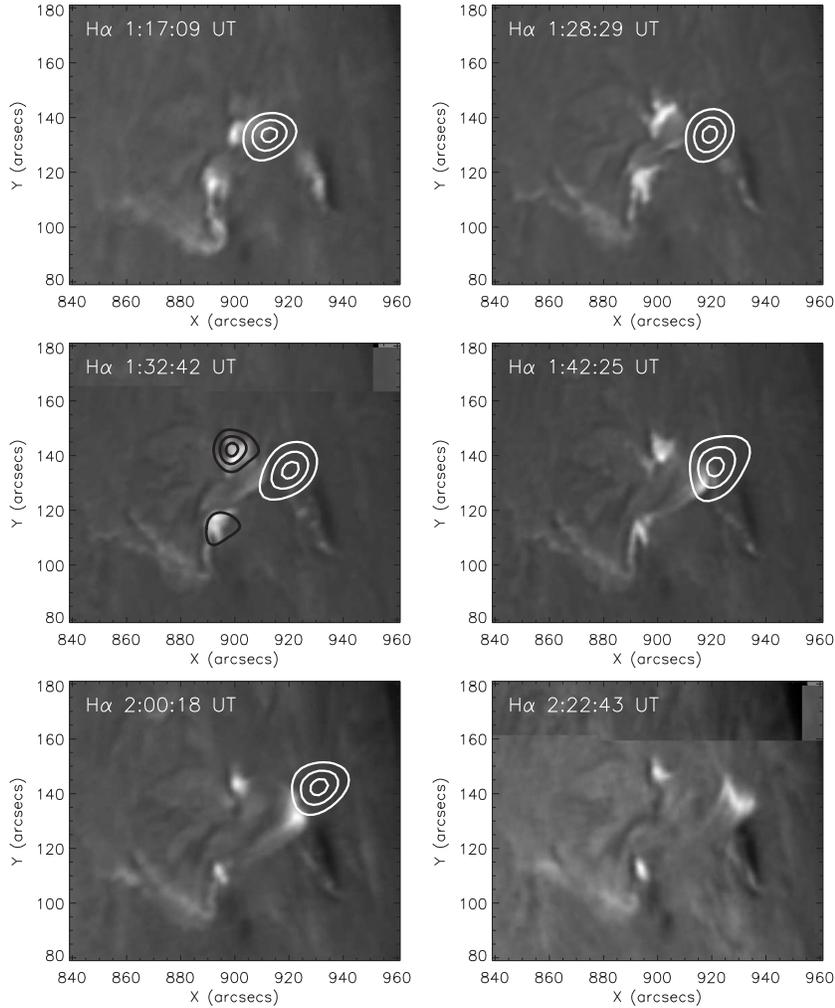}
\caption{Sequence of H$\alpha$ images overlayed by RHESSI X-ray images which are closest in time. 
White contours are 10--15~keV, black contours (only in the third panel) are 50--100~keV.  
The contour levels are 60, 80 and 95\% of the peak flux of each image.}
\label{ha_rh} 
\end {figure}

\begin{figure}
\centering
\includegraphics[width=11cm]{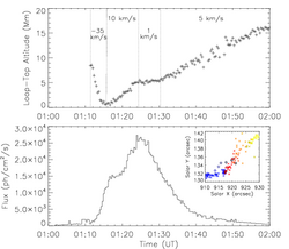}
\caption{$Top~panel$: Evolution of the altitude of the RHESSI LT source observed in the 10--15~keV 
energy band. (Note that the zero level is arbitrarily chosen.) The mean velocities
derived by linear fits to the altitude data in certain time intervals (indicated by vertical lines) are 
annotated in the figure. $Bottom~panel$: Light curve of the LT source in the 10--15~keV band. 
The integration time for each image used to derive the centroids and the flux of the LT source
was 20~s. The $inset$ shows the RHESSI LT source centroids with the time evolution color coded
from $dark~(blue)~to~bright~(yellow)$. $Diamonds$ indicate the centroids during the altitude decrease, $crosses$
are for the rising motion.}
\label{rh_kin} 
\end {figure}

Figure \ref{rh_lc} shows hard X-ray light curves in the 10--25, 25--50, 50--100 and 100--300~keV 
energy bands observed by the RHESSI instrument (Lin et al., 2002). The RHESSI observations 
cover the total impulsive phase of the flare but are contaminated by X-rays from a particle event, 
i.e.,\ the RHESSI detectors were hit by high-energy particles trapped in the Earth's radiation belts.
The X-ray fluxes above about 30~keV are strongly affected by the particle event and therefore the time
histories $\gtrsim$~30~keV have to be interpreted with caution since a substantial fraction of the signal is 
of non-solar origin.

In the hard energy channels, 100--300 
keV, 50--100 keV, and 25--50 keV, the count rates peak around 01:18 UT 
and the maxima coincide with the time of the first maximum observed in 
the GOES profiles (refer to Figure \ref{goes_halpha}a). Further, we observed another peak at 01:21 UT
equal to the one at 01:18 UT.  The first one was broad and
lasted for about 4~min whereas the second peak was narrower. These
peaks occur nearly simultaneously at three high energy bands.
However, in the lower energy channels (probably dominated by thermal bremsstrahlung emission), 
i.e., 10--25 keV, and in the GOES profiles, the count rates maximized after 01:24~UT. 

Figure \ref{ha_rh} shows a sequence of H$\alpha$ images overlayed with RHESSI X-ray images. 
The RHESSI images have been reconstructed with the Clean algorithm with
the natural weighting scheme using front detector segments 3 to 8 in the energy bands 10--15~keV and
50--100~keV (Hurford et al., 2002). At high energies we were able to reconstruct RHESSI images only 
at the last X-ray peak around 01:33 UT (cf. Figure \ref{rh_lc}). At this time, two footpoints are 
observed in hard X-rays which are roughly co-spatial with the two strong footpoints observed in H$\alpha$ 
(see Figure \ref{ha_rh}, third panel). Due to the strong contribution from the particle event, it was not
possible to reconstruct RHESSI footpoint images at earlier times. The signal 
(solar) to ``noise'' (particle event) ratio
is as low as 1:50 during these times (Gordon Hurford, private communication). However, in H$\alpha$ 
two strong footpoints are observed during the whole impulsive phase which separate from each other
in the course of time.

At low energies ($\lesssim$~30~keV) the X-ray emission is concentrated in a distinct looptop (LT) 
source located above the H$\alpha$ LT source at each instant, 
as it is expected since the H$\alpha$ LT source is generally considered to be the result of cooling (and also shrinking) of the 
hot X-ray emitting flare loops (cf.\ Vr\v{s}nak et al., 2006a, and references therein).
The RHESSI LT source can be observed as early at 01:02:20~UT when the X-ray level is still very low 
until the end of RHESSI observations at about 02:01~UT when the
spacecraft again entered the Earth's shadow. 

Figure \ref{rh_kin} shows the altitude evolution 
of the RHESSI LT source which was derived by measuring the centroids of the LT emission in the
10--15~keV energy band along its main axis of motion. 
The angular resolution of the RHESSI images which were reconstructed with front 
detector segments 3 to 8 is $\sim$8$''$. However, we stress that the 
determination of the emission centroids of the sources may be as good as 1$''$, 
depending on the count statistics and source complexity (Hurford et al., 2002).
From 01:16~UT up to the end of the RHESSI 
observations at 02:01~UT the RHESSI LT source rises. In the classical ``CSHKP'' model of eruptive 
(two-ribbon) flares this rising motion as well as the footpoint separation reflects the progression of 
magnetic reconnection during which field lines rooted successively further apart from the magnetic 
inversion line reconnect. The highest upward growth is observed between 01:16 and 01:24~UT 
with a mean velocity of 10~km~s$^{-1}$. Afterwards the loop growth slowed down to 1~km~s$^{-1}$ 
between 01:24 and 01:30~UT, and rose again with a mean value of 6~km~s$^{-1}$ 
during the interval 01:30--02:00~UT. We note that these are comparatively
low velocities for the growth of the flare loop system in an X-class flare (for instance,
in the homologous X-class flare at 09:45~UT the velocities of the LT growth are about 
a factor~2--3 larger, see Fig.~6 in Veronig et al., 2006 and Table~1 in Liu et al., 2004).  
The low velocities are most probably due to the fact that the flare loop 
footpoints are rooted in sunspots (cf.\ the inset in Figure \ref{mdi} and the last panel in Figure \ref{ha_rh}), i.e.,\ 
regions of very strong magnetic fields~$B$: although the loop growth and ribbon expansion
velocities~$v$ are small, the locally reconnected magnetic flux $\sim {\bf v} \times {\bf B}$ can be very large.

Between $\sim$01:11 and 01:16~UT, i.e.\ at the beginning of the impulsive phase, the height of the LT source 
apparently decreases with a mean velocity of 35~km~s$^{-1}$ (cf. Figure \ref{rh_kin}). Such a decrease of the LT altitude 
in the early impulsive phase has been recently reported for several 
flares observed by RHESSI (Krucker, Hurford, and Lin, 2003; Sui and Holman, 2003; Sui, Holman, and Dennis, 2004;
Liu et al., 2004; Ji et al., 2006; Veronig et al., 2006) as well as in soft X-rays by GOES12/SXI, in TRACE EUV
and in microwaves (Li and Gan, 2005, 2006; Veronig et al., 2006), and is assumed to be closely related
to the establishment of the magnetic reconnection process powering the flare (cf. the discussion in Sui, Holman, 
and Dennis, 2004).

\subsection{EIT Observations}

\begin{figure}
\centering
\includegraphics[width=11cm]{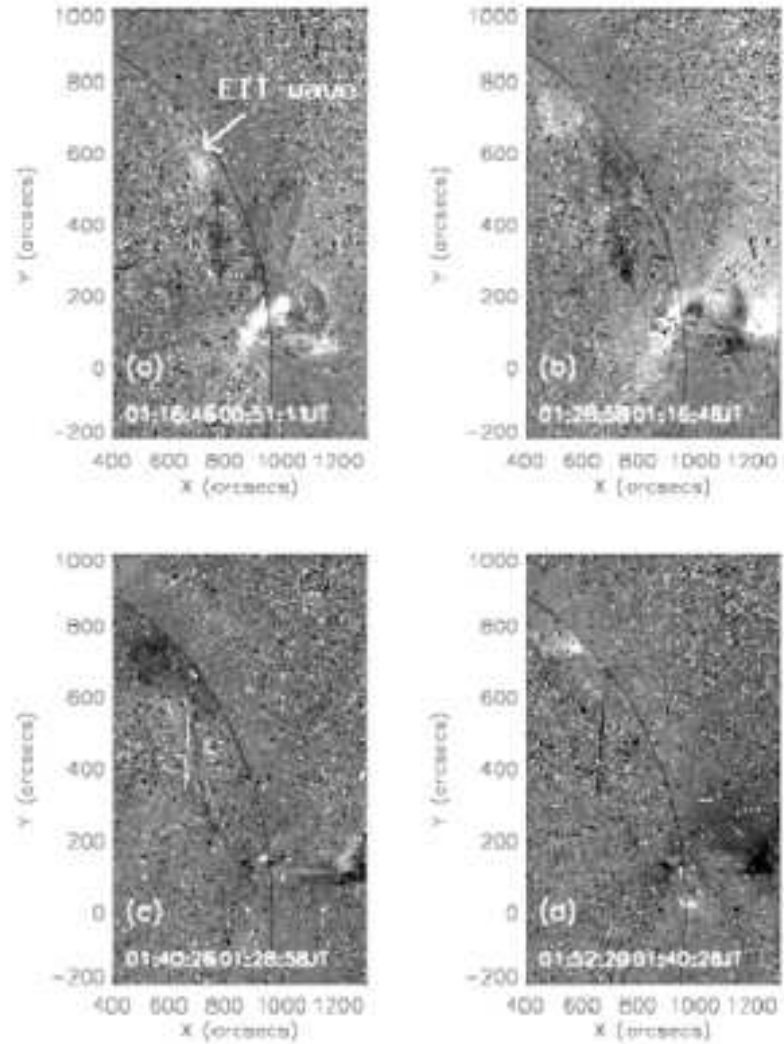}
\caption{Running difference images derived from SOHO/EIT observations in the 195~{\AA} line showing the sequence of activities from pre-flare to peak of the event.}
\label{eit}
\end{figure}                                                                    

Figure \ref{eit} shows the EUV images of the flare and its environment observed at
195~{\AA} with EIT onboard the SOHO spacecraft (Delaboudiniere et al., 1995). These running difference
images reveal the sequence of activities from the beginning to the declining phase
of the flare. During the rising phase (cf. Figure \ref{eit}a) the eruption is evident
where a patchy pattern of high intensity is observed on the disk, close to the active
region. Moreover, two looptops are seen  
outside the limb towards the west side of the active region. 
The southern looptop is brighter than the one located just above, 
towards the north. 

Figure \ref{eit}b, which corresponds to the impulsive phase of the flare,  shows
eruption of plasma from the looptop. The intense emission observed close to the flare site
in the previous image continues here. Further, the emitting region at the top of the loop
system shows complex structure, suggesting a crucial stage at the time 
of the reconnection process, i.e., prior to the restructuring of the 
magnetic field after the reconnection followed by the ejection of
plasma and field. The images recorded at the later time intervals (Figures \ref{eit}c and \ref{eit}d)
show relaxation of field at the flare region and the intensity of
the emission declines heavily. These results are consistent with RHESSI and H$\alpha$ findings.

The EIT difference images also reveal the propagation of an EIT wave
associated with the flare event (Thompson et al., 1999). In Figure \ref{eit}a, the EIT wave is visible on the disk 
moving towards the north-east of the flaring region. The wave propagates further towards the north direction
in the next image (Figure \ref{eit}b). 

\subsection{Radio Spectra by HiRAS and WAVES}

\begin{figure}
\epsfig{file=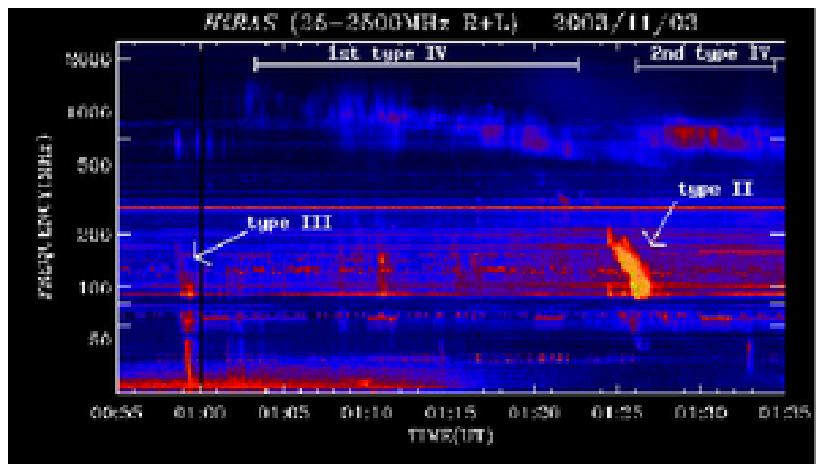, width=12cm, height=8cm}
\caption{The dynamic radio spectrum of the event from 25 to 2500 MHz observed 
by the HiRAS spectrograph.}
\label{hiras}
\end{figure}                                                                    

\begin{figure}
\epsfig{file=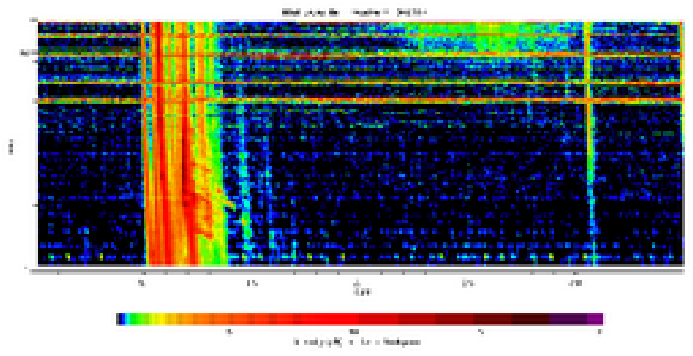, width=13cm, height=8cm}
\caption{WAVES/Wind spectrum observed on 3 November 2003 during 00:30 to 03:30 UT from 1 to 12 MHz.}
\label{wind}
\end{figure}                                                                    

The HiRAS spectrograph operated by the NICT, Japan, observed 
significant activities at a wide range of frequencies between 30 and
2000 MHz in association with this LDE. 
Figure \ref{hiras} shows the HiRAS spectrum.  The radio emission
starts at 00:59 UT with a bunch of type III bursts from 1000 MHz to the lowest
observed frequencies, and is followed by two 
type IV continua one after the other between 01:02 and 01:35 UT.
Both broad-band type IV continua (i.e., in the 
frequency-range 400$-$1500 MHz) showed a slow drift 
toward the low frequency side.  The first type IV continuum starts
at a frequency of $\sim$1500 MHz and its width extended
down to about 400 MHz between 01:02 and 01:22~UT. However, the second continuum was
mostly confined between 1000 and 500 MHz and is observed during 01:26$-$01:35 UT.
The slow drifts observed in both Type IVs suggest the expansion of loops
filled with non-thermal particles. The typical heliocentric height of
these bursts can be at $\sim$1.1~R$_\odot$. The drift rate of the first
continuum suggests a typical expansion rate of about 40 to
60 km s$^{-1}$ ($\sim$01:02 to 01:22 UT). However, during this time interval, between 01:12 and 
01:16 UT, the LT source as seen by RHESSI (cf. Figure \ref{rh_kin}) shows a downward
altitude shift. It is likely that the expansion of the loop system
caused a major reconfiguration of field lines.
The second continuum seems to represent a lower expansion rate of $\le$ 40
km s$^{-1}$ (at time $>$~01:25 UT).

Additionally, the spectrum of the first continuum showed vertical structures 
of brighter intensity at random time intervals. 
However, between 01:15 to 01:21 UT, these vertical structures formed two patches of high intensity
with a frequency drift from 1000 MHz to 600 MHz. These patches 
of intense flux density represent rapid acceleration of
electrons within a range of defined heights (i.e.,
$\sim$2 -- 4 $\times$ 10$^4$ km), and they are likely to 
be associated with the particle acceleration along the 
current sheet during the reconnection 
process. 

An intense type II burst is observed between 01:24 and 01:28 UT, in 
the frequency range of $\sim$200$-$80 MHz. The intensity
of this type II is low at the higher frequency side. 
However, a careful examination reveals the fundamental emission of
this burst running parallel to the above intense burst at
a lower-frequency range. As it has been reported in
the literature (e.g.\ Kundu, 1965)
the typical strength of the second harmonic seems to be more
intense. 
Using the Saito coronal density model (Saito, 1970), 
we estimate the heliocentric height of the type~II to be about 1.2~R$_\odot$. 
It may be noted that the second harmonic of
the type II has two regions: a fairly slow drift up to
about $\sim$01:25 UT, followed by the onset of a sharp--fast drift structure.
The slow--drift part seems to agree with a speed of about
700 to 800 km s$^{-1}$, whereas the fast--drifting structure is
consistent with a speed of about 1200 km s$^{-1}$. 
The typical height of the type II emission is in good
agreement with the EIT wave discussed above (e.g., Figure \ref{eit}).
The occurrence of the very intense fast--drifting type~II suggests
the ejection of mass, and its starting frequency, $\sim$200 MHz,
confirms the ejection at a height above the 
first type IV continuum. In other words, the restructuring
of magnetic field lines and the energy release of the
flare/CME are revealed by the type II and type IV bursts as 
well as the structure embedded within them. 

Here it is interesting to note that the EIT wave was observed a few minutes prior to 
the start of type II burst (cf. section 2.4 and Figure \ref{eit}). The association between
the EIT wave and the coronal shock wave (i.e. type II) suggests that both wave phenomena
are probably different signatures of same coronal disturbance initiated by
the flare or CME (see Klassen et al., 2000).

Just after the type II burst, a second slow moving type IV continuum starts 
at 1000 MHz and drifts to lower frequencies between 01:26 to 01:35 UT.
On the whole, this type IV continuum is brighter than the first one discussed
above. The radio emission from this continuum becomes very intense 
between 01:26 to 01:33 UT.
We also find that, in the time interval of 01:26 to 01:29 UT, there are two bands of continuum bursts
that lie one over the other on the frequency scale, suggesting the 
presence of two radio--emitting structures containing high-energy 
accelerated electrons.

The type III bursts observed at 00:59 UT in the HiRAS spectrum
continued in the whole frequency range of 1--14 MHz of the WAVES 
experiment on the Wind spacecraft (Figure \ref{wind}). But, on the WAVES spectrum, 
multiple intense bursts were observed  between 01:00 and 01:30 UT. 
Also, finger like structures were visible after the first maximum 
of the event
around 01:24 UT in the frequency range of 1--2 MHz, indicating 
the production of accelerated particles after the first stage
of reconnection.  In the high-frequency part of the WAVES spectrum
(i.e., 9--14 MHz), an intense patch of emission was observed 
between 02 and 03 UT. The emission height of this radio burst suggests
that it is likely to be the continuation of the slow drifting type IV continua observed
in HiRAS spectra, and hence is associated with the expansion of an arcade of loops
to a height $>$ 2 R$_\odot$.

\subsection{Coronal Mass Ejection -- White-light and IPS Images}

\begin{figure}
\epsfig{file=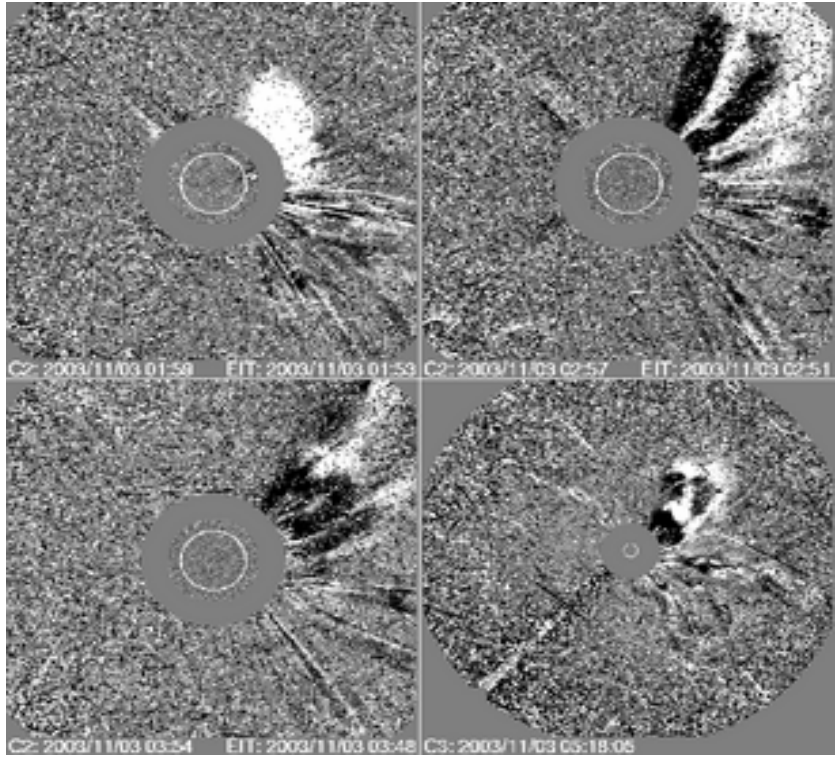, width=11cm, height=10cm}
\caption{Running difference images derived from LASCO C2 and C3 showing
the propagation of the CME associated with the flare.} 
\label{cme}
\end{figure}

\begin{figure}
\hbox{
\epsfig{file=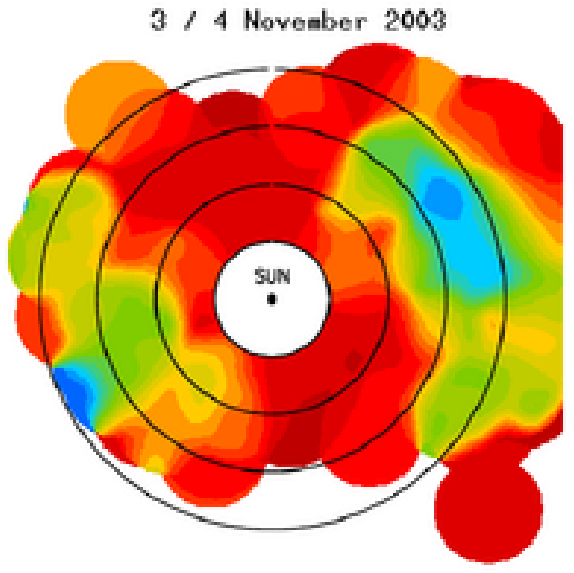, width=5.5cm, height=5.5cm}
\epsfig{file=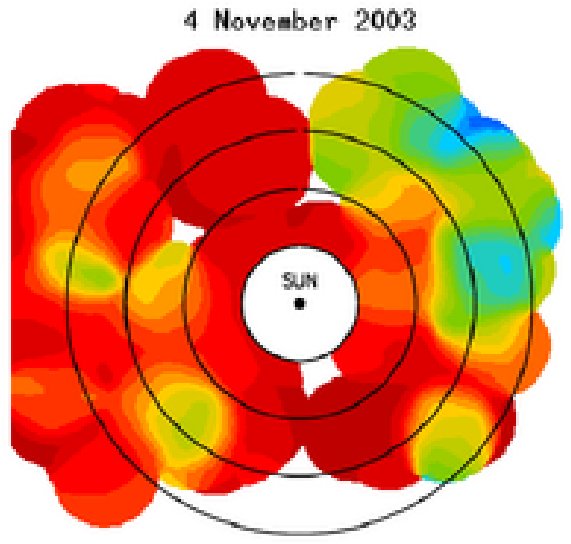, width=5.5cm, height=5.5cm}
\epsfig{file=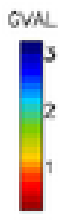, width=0.8cm, height=5.5cm}
}
\caption{Ooty scintillation images for 2003 November 3 and 4 measurements. The 
concentric circles are of radii 50, 100, 150 and 200 R$_\odot$. The red color
code indicate the background solar wind.}
\label{ips}
\end{figure}                                                                    

The Large-Angle Spectroscopic Coronagraph (LASCO) on board SOHO 
images the white-light corona from 2$-$30 R$_\odot$ 
(Brueckner et al., 1995). The C2 coronagraph covers a field of 
view of 2$-$6 R$_\odot$ while the C3 coronagraph images the corona 
from 4$-$30 R$_\odot$. A CME, with a fairly wide bright loop front with 
cavity and gusty outflow moving towards the north-west side of the Sun, 
was associated with this flare event, (cf. Figure \ref{cme}). The CME
onset was first seen in the LASCO C2 field of view at 01:59 UT. 
The angular width and 
position angle of the CME were 65$^{\circ}$ and 304$^{\circ}$, respectively.
At first, the CME appeared in C2 at a radial distance of $\sim$4 R$_\odot$ 
and it could be followed by C3 up to $\sim$20 R$_\odot$.  
In the C2 field of view, the CME appears to 
have a bright asymmetrical loop structure. The successive 
images show the fast movement of the CME with time. The CME maintains its 
loop-like structure till 02:57~UT when its leading edge goes 
beyond the C2 field of view. The CME first appears in the C3 field of view at 02:22~UT and can
be seen till $\sim$06:42~UT when its leading edge becomes very faint.
The height-time plot available at the SOHO LASCO-CME catalogue\footnote{http://cdaw.gsfc.nasa.gov/CME$\_$list/}
shows that the mean propagation speed of the CME in the LASCO field of view
is about 827~km~s$^{-1}$. However, the first two points in the height-time
plot of the CME observed by C2 indicate that the CME was faster at the beginning with
a speed of about 1000 km s$^{-1}$. A second$-$order polynomial fit to 
the height$-$time data indicates a deceleration of $\sim$28~m~s$^{-2}$ 
in the CME propagation at heliocentric heights less than 30~R$_\odot$. 

Figure \ref{ips} shows interplanetary scintillation (IPS) images of the
CME observed with the Ooty Radio Telescope, Radio Astronomy Centre.
The IPS technique measures the turbulent plasma at the front of
the moving CME. The above images are projected on the sky plane
and with these measurements the CME could be followed outside the
LASCO field of view up to about 1 AU. In this heliocentric
distance range the speed of the CME declined approximately as a
power law of  R$^{-0.4}$, suggesting that the magnetic field 
associated with the CME supported the propagation to a fairly large
distance of about 100 R$_\odot$ (Manoharan et al., 2001; Manoharan, 2006).
The CME deceleration, mostly effected by the typical interaction
with the background solar wind flow (see, e.g., Vr\v{s}nak et al., 2004, and references therein), 
seems to be rather slow. This indicates that 
additional energy is provided by the internal energy of the CME
(i.e., possibly magnetic field expansion).
The IPS images indicate that the CME associated shock propagates into the
inner heliosphere to large distances from the Sun, while the IPS
intensity indicates the strength of compression region between the shock
and the driver gas, i.e., CME. Further, the speed profile
and its comparison with IPS measurements made on a number of
CMEs (Manoharan 2006), suggest an additional energy supplied
by the CME to the propagation.  

\section{Discussion}

The 2B/X2.7 flare that occurred on November 3, 2003 has been observed
with a set of instruments, which cover a wide range of wavelengths
and energy bands. 

The multi-wavelength observations discussed in the previous section provide us some important
features and sequence of activities before the rising phase of the event that need more attention.
The HiRAS spectrograph shows the type III radio burst at $\sim$00:59 UT, significantly earlier than the
flare signatures in soft X-rays and H$\alpha$. The type III started at 1000 MHz, very low in the corona
and is followed by type IV that started at even lower coronal heights. The frequency drift of this type IV 
shows the bulging of the magnetic field system out into the corona. In HXR, a 10-15 keV
LT source is observed in low corona as early as $\sim$01:02 UT. 
We interpret these observations as evidence for the first phase of a breakout, when reconnection in the corona
transfers overlying flux to adjoining regions, thus weakening the constraining magnetic configuration
(Antiochos, DeVore, and Klimchuk 1999). 
However, in the GOES profile the first peak
is seen at 01:17 UT, which we attribute to the filling of thermal particles in the flare loops in response to the
phase of the breakout process. The H$\alpha$ images during this interval (i.e., between
the type III and the first peak in GOES profile) show multiple brightenings in the chromosphere 
and a spray that shows evaporation of plasma from the chromosphere towards
the north-east side of the active region.

During the whole impulsive and gradual phase, the H$\alpha$ observations clearly show 
two strong well$-$defined footpoints which separate from each other in the course of time. 
After the flare maximum, a loop system is observed which connects these two footpoint sources. 
With the evolution of the flare, the loop system grows and intense emission is produced 
at the looptop. The emission at the two footpoints in H$\alpha$ is co-spatial with the 
RHESSI image in the 50--100 keV energy band.
On the other hand, the evolution of the looptop source is also evident in RHESSI
images in the 10--15~keV energy band, and the RHESSI LT source is always
located above the LT source in H$\alpha$. 
The growth of the loop system has also produced signatures
in radio observations obtained from the HiRAS spectrograph between 1000--400 MHz frequency range
as two broad band type IV radio bursts drifting towards the low frequencies one after the other. 
An intense type II burst, emission most probably at the second harmonic, occurred in between these type IV 
continua but at a lower frequency 
range ($\sim$200$-$80 MHz). The type II shows the onset of the CME and the associated coronal shock wave.
EIT observations indicate intense brightening at the activity site as well as large loops associated
with the active region. The H$\alpha$, RHESSI, and EIT observations along with the associated radio emission observed
over a wide range of frequencies indicate the link between the energy release in the corona, downward to the
chromosphere and upward in the heliosphere.

In the present analysis we have discussed the evolution of a flare and
associated eruption that took place in a complex multipolar magnetic topology.
The H$\alpha$ and RHESSI observations clearly demonstrate the growth of a low
lying loop system with the hard X-ray source at the top of the loops. 
With the evolution of the looptop source, new hot loops are formed with ever 
increasing altitude while the
lower ones cool down and begin to be visible in H$\alpha$ in the later stages.
The EIT images show the magnetic field structure associated with the active region
at larger heights in the corona. 
The CME structure observed in IPS images and their speed profiles 
suggest that the CME also carried sufficient magnetic energy.
In summary, we find that the flare evolution is consistent with the standard CSHKP model of
solar flares. The observations also contain evidence for breakout reconnection before the flare
phase.

\begin{acknowledgements}
It is a great pleasure to thank Ram Sagar for 
his support and encouragement in the present study.
The authors thank the observing and engineering staff of Radio Astronomy Centre
in making the IPS observations. We also acknowledge the 
technical staff of ARIES Solar Tower Telescope for their assistance in 
maintaining and making regular solar observations. 
We acknowledge RHESSI and SOHO for their open data policy. RHESSI is a NASA's small 
explorer mission. SOHO is a project of an 
international collaboration between ESA and NASA. The authors would
like to thank for the excellent LASCO-CME catalog, which includes the 
supportive data. The CME catalog is generated
and maintained by the Center of Solar Physics and Space Weather, Catholic
University of America, in cooperation with the Naval Research Laboratory
and NASA. We also acknowledge the team of Wind/WAVES and HiRAS for radio 
dynamic spectra. We thank Gordon Hurford for valuable discussions on RHESSI
data analysis. The constructive comments and suggestions from an 
anonymous referee are sincerely acknowledged. 
We thank Judy Karpen for insightful comments and valuable discussions that significantly
improved the scientific content of the paper. 
\end{acknowledgements}

\end{article}
\end{document}